\documentclass[3p,times]{elsarticle}%
\journal{***}

\usepackage{amssymb}
\usepackage{amsmath}
\usepackage{amsthm}
\usepackage{graphicx}% Include figure files
\usepackage{bm}% bold math
\usepackage{hyperref}% add hypertext capabilities
\usepackage{xcolor}
\usepackage{float}
\usepackage{doi}
\usepackage{natbib}
\usepackage{subfig}
\biboptions{sort&compress}

\begin{document}
	
\begin{frontmatter}

		\title{Community detection and portfolio optimization}
		\author[add1]{Longfeng Zhao\corref{cor1}}
		\ead{zlfccnu@nwpu.edu.cn}
		\ead{zlfccnu@bu.edu}
		\cortext[cor1]{Corresponding author}
		\author[add2]{Chao Wang}
		\author[add3]{Gang-Jin Wang}
		\author[add4]{H. Eugene Stanley}
		\author[add1]{Lin Chen}
		\address[add1]{School of Management, Northwestern Polytechnical University, Xi'an, China}
		\address[add2]{Research Base of Beijing Modern Manufacturing Development, College of Economics and Management, Beijing University of Technology, Beijing, China}
		\address[add3]{Business School and Center for Finance and Investment Management, Hunan University, Changsha 410082, China}
		\address[add4]{Center for Polymer Studies and Department
				of Physics, Boston University, Boston, Massachusetts
				02215, USA} 
\begin{abstract}

Community detection methods can be used to explore the structure of
complex systems. The well-known modular configurations in complex
financial systems indicate the existence of community structures. Here
we analyze the community properties of correlation-based networks in
worldwide stock markets and use community information to construct
portfolios. Portfolios constructed using community detection methods
perform well. Our results can be used as new portfolio optimization 
and risk management tools.
		
\end{abstract}
		
\begin{keyword}

stock market \sep correlation-based network \sep community detection \sep
			portfolio optimization

\end{keyword}
	
\end{frontmatter}

\section{Introduction}

\noindent
Community detection is an an effective tool for exploring the modular
structures in complex networks \cite{Fortunato2016}, because community
properties indicate how complex systems are organized. Since Newman
\cite{Newman2004a} proposed the first formal definition of the modular
structure of complex networks, the community detection task has become a
major focus in network theory, and numerous community detection
algorithms have been proposed
\cite{Fortunato2010,Han2016a,Han2017}. Among the many real-world
modularized complex systems, financial markets have become a
particularly attractive subject, but the coexistence of noisy and
structured patterns make financial systems difficult to deal with. To reduce
the influence of noise, many correlation-based methods have been
proposed to extract structured information from the complex correlations
among financial assets, including the threshold method, the minimum
spanning tree (MST) \cite{Mantegna1999a}, the planar maximumly filtered
graph (PMFG) \cite{Tumminello2005}, among others
\cite{Kenett2010,csm,Gao2015,Gan2015,WangGJ2017a,WangGJ2017,WangGJ2016}.

These correlation-based networks use complex cross-correlation matrices
to determine the noise-free structure of complex financial markets.
Because correlation-based network methods differ from the
eigenvector-based methods that use variance contributions from only a
few dimensions, they can transform dense correlation matrices into
sparse representations, i.e., complex networks. These correlation-based
network methods are popular because the noise-free information they
produce can be used to analyze interactions among financial assets
\cite{Marti}.
 
Much recent research has used the structured information provided by
correlation-based network to optimize portfolio selection procedure
\cite{Pozzi2013,Peralta2016,Zhao2017b,Zhao2018a}.  In
particular, some compound centrality measures using centrality ranking have
been proposed to quantify the importance of financial assets. From a
complex network perspective, the modular structures in financial
networks link correlation-based networks and community structures, and
we thus can assume that the community structures in correlation-based
networks can also provide useful information about risk diversification.

We here use a community detection algorithm to analyze the
correlation-based networks in stock markets. We construct a strand of PMFG networks
for the US, UK, and Chinese stock markets \cite{Tumminello2005}, the
community structures of which are readily available. We then use this
community information to construct portfolios of stocks from
different communities. The good performance of these portfolios,
indicates the usefulness of the community structures.

Section \ref{data} describes the data and methodology we
use. Section~\ref{results} presents our results, including a topology
analysis of stock markets and its application to portfolio
optimization. Section~\ref{conclusion} presents our conclusions.

\section{Data and methodology\label{data}}

\subsection{Data}

\noindent
Our data sets are of the daily adjusted returns of the constitute stocks
of three major indexes: the S\&P 500 (US), the FTSE 350 (UK), and the
SSE 380 (China). To accurately estimate cross-correlations among stocks,
we discard the stocks where the sample size is small.  This leaves us
with 401, 264, and 295 stocks for the three markets, respectively. Each
stock in the S\&P 500 index has 4025 daily returns from 4 January 1999
to 31 December 2014. Each in the FTSE 350 index has 3000 daily returns
from 10 October 2005 to 26 April 2017. Each in SSE 380 index has 2700
daily returns from 21 May 2004 to 19 November 2014.

\subsection{Community detection of the correlation-based networks}

\noindent
Before we introduce the community detection algorithm, we need to construct 
correlation-based networks from the return time series. We define logarithm 
return $r_{i}(t)$ to be
\begin{equation}
r_{i}(t)=\mathrm{ln}(p_{i}(t+1))-\mathrm{ln}(p_{i}(t)).
\end{equation}
\noindent
Here $p_{i}(t)$ is the closure price of stock $i$ at time $t$. We then
use the past return records inside a moving window of length $\Delta$ to
calculate the Pearson cross-correlation coefficients between any pair of
return time series at time $t$.  We obtain an $N\times N$ matrix
$\mathrm{\textbf{C}}^{t,\Delta}$ at time $t$ with an estimation window
of $\Delta$ days, and $N$ is the number of stocks. We select the moving
window widths to ensure the non-singularity of the cross-correlation
matrix ($\Delta\geq N$) with $\Delta=500$ days for the S\&P 500 and
$\Delta=300$ days for both the FTSE 350 and the SSE 380. We then shift
the moving window with 25-day step and obtain a strand of correlation
matrices for the three markets. We now have 142, 109, and 97 correlation
matrices for the S\&P 500, FTSE 350, and SSE 380 indices, respectively.

We then employ the planar maximally filtered graph (PMFG) method to remove
redundant information induced by the dense cross-correlation
matrices \cite{Tumminello2005}, and we construct sparse networks from the
cross-correlation matrices $\mathrm{\textbf{C}^{t,\Delta}}$. A planar
graph $G^{t,\Delta}$ is formed with $N_e=3(N-2)$ edges in a system with
$N$ stocks. Reference~\cite{Tumminello2005} indicates that the PMFG
keeps the hierarchical organization of the MST and also induces
cliques. The PMFG network can discriminate between important information
and noise. Previous studies have analyzed the modular structure of the
PMFG network for both indices and stocks
\cite{Song2011,Zhao2015b,Zhao2018a}, and we here use the
infomap algorithm \cite{Rosvall2008} to detect the community structures
in the PMFG networks for the three markets. A comparative analysis
indicates that the infomap algorithm is the best-performing community
detection algorithm\cite{Lancichinetti2009}. The community structure 
given by the infomap
algorithm provides a comprehensive description of the macrostructure of
financial markets. Thus a natural generalization is to use community
information when optimizing portfolio design.

\section{Results and applications}
\label{results}
\subsection{Community structure}

\begin{figure}
	\centering
	\includegraphics[scale=0.6]{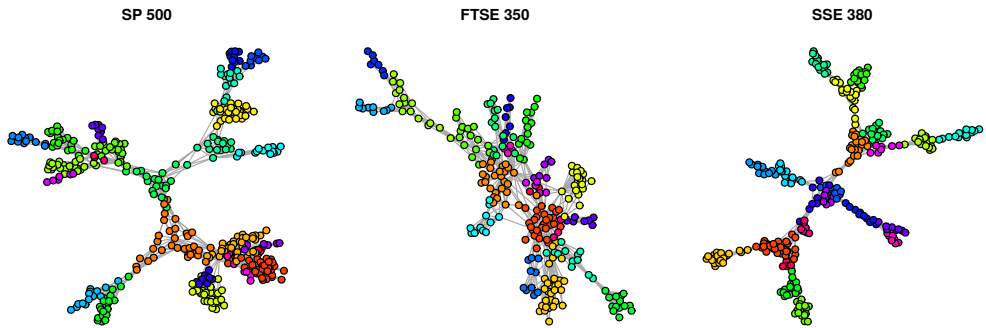}
\caption{Three typical networks for three stock markets. Different color
  represents different communities. }
	\label{fig:pmfg}
\end{figure}

\noindent Figure~\ref{fig:pmfg} shows three PMFG networks for the three
stock markets. We use the infomap algorithm to obtain the community
structures, which are shown in different colors. The PMFG networks each
have a distinctive community structure, and the appearance of hub stocks
indicates that stock market structures are heterogeneous.

\begin{figure}
	\centering
	\includegraphics[scale=0.6]{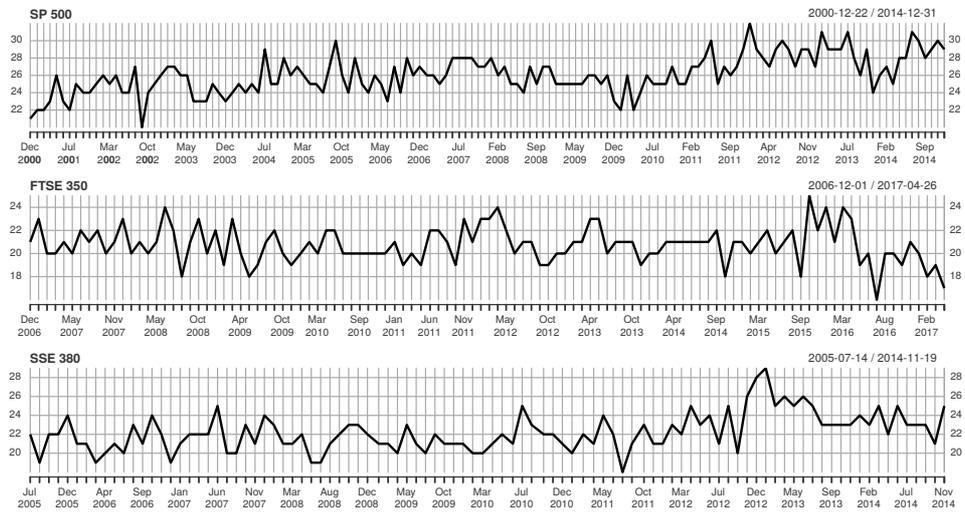}
\caption{The number of communities for three markets over time.}
	\label{fig:communityNum}
\end{figure}

Figure~\ref{fig:communityNum} shows the number of communities detected
in the three markets. The mean number of communities is 26 for the S\&P
500 index, 21 for the FTSE 350 index, and 22 for the SSE 380 index. The community numbers
usually fluctuate around its mean value. Note that the average number of
communities can indicate an approaching crisis period, such as the
sub-prime crisis beginning in September 2008 and the European debt
crisis at the end of 2009. Because fluctuations in the number of
communities makes drawing useful conclusions about market states
difficult, however, we can analyze the modularity $Q$ of the PMFG for the
three markets, which is defined
\begin{equation}
Q=\frac{1}{2m}\sum\limits_{vw}\left[ a_{vw}-\frac{k_v
    k_w}{2m}\right]\delta(c_v,c_w), 
\label{modularityEquation}
\end{equation}
where $a_{vw}$ is the entry of the adjacency matrix of the PMFG, $k_v$
the degree of node $v$, $m$ the number of total edges of PMFG, and $c_v$
the community membership of node $v$.  The modularity $Q$ quantifies with a
value between 0 to 1 the level of modular structure in the
correlation-based network.  In every PMFG at any given time, one
community partition will give a modularity value that indicates the
degree of the modular structure of the market.

\begin{figure}
\centering
\includegraphics[scale=0.6]{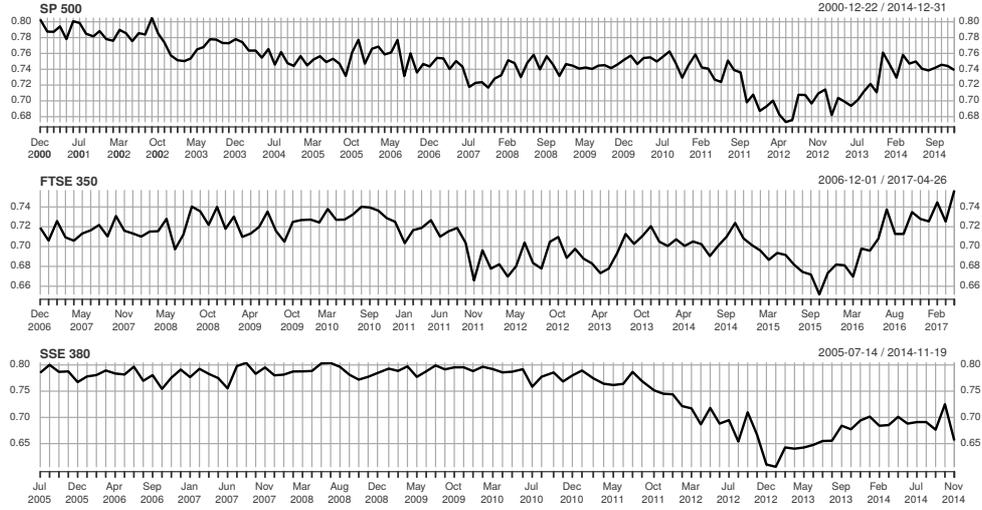}
\caption{The modularity of three markets over time.}
\label{fig:modularity}
\end{figure}

Figure~\ref{fig:modularity} shows the evolution of the modularity for
the three markets, which decreased rapidly during the European debt
crisis. Note that the modularity prior to 2011 in the Chinese market is
very high, indicating that the Chinese market is highly structured.

\begin{figure}
	\centering
	\includegraphics[width=\textwidth]{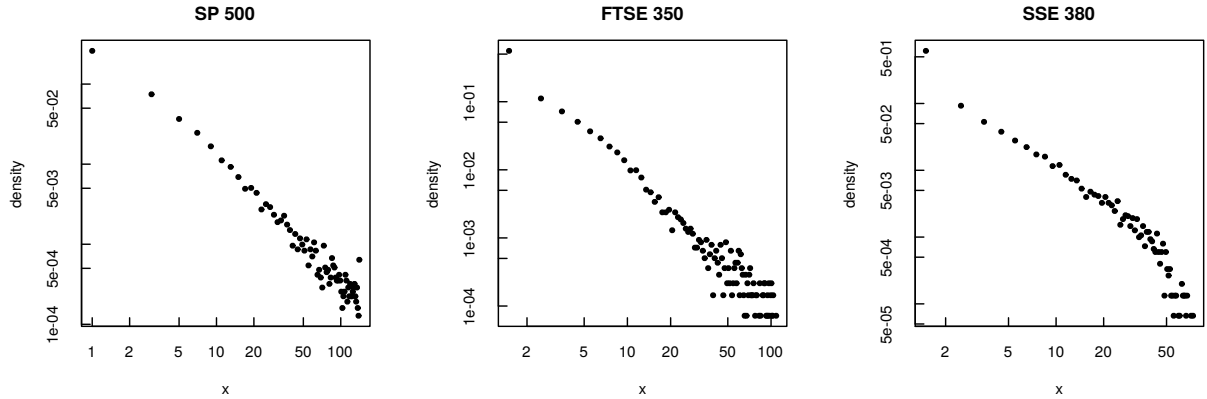}
	\caption{The co-occurrence distribution for three markets.}
	\label{fig:occurrencedist}
\end{figure}

We also use these time-evolving PMFGs to analyze the stability of the
connections in the three markets. We checked the co-occurrence time $t$
of stock pairs within the same community during the whole observation period. 
We use the co-occurrence time to evaluate the stability of the
community structure. An extremely stable community structure produces a
double peak distribution with co-occurrence time $t=0$ and $t=T$. When
$t=0$ the two stocks never stay in the same community. When $t=T$ the two
stocks are always in the same community. Figure~\ref{fig:occurrencedist} shows
extremely heterogeneous distributions, indicating that the
community structures of the three markets are extremely unstable over
time. Note that the distributions of the
co-occurrence times are power laws in the three markets. Although a few
stocks belong to the same community for a long period of time, market
fluctuations cause the stay of most stocks in the same community to be
very short.

Thus these community structures produce information about the structure
evolution in the three markets. We will now use this structured
community information to optimize the portfolio construction procedure.

\subsection{Portfolio optimization}

\subsubsection{Mean-variance portfolio optimization}

\noindent We first employ the community structure to 
improve the performance of portfolio optimization under the Markowitz portfolio optimization
framework\cite{Markowitz1952}. Much research has focused on establishing
connections between correlation-based networks and portfolio
optimization problems
\cite{Onnela2003b,Tola2008235,Pozzi2013,Peralta2016,Zhao2017b,Zhao2018a}.
We now give an brief introduction about the Markowitz portfolio theory.
Using Markowitz portfolio theory we examine a portfolio of $m$ stocks
with return $r_i,i=1\ldots m$. The return $\Pi(t)$ of the portfolio is
\begin{align}
\Pi(t)=\sum\limits_{i=1}^{m}\omega_{i}r_{i}(t),
\end{align}
where $\omega_{i}$ is the investment weight of stock $i$, and
$\omega_{i}$ is normalized such that
$\sum\limits_{i=1}^{m}\omega_{i}=1$. We quantify portfolio risk using
the variance of portfolio return
\begin{align}
\Omega=\sum\limits_{i=1}^{m}\sum\limits_{j=1}^{m}
\omega_{i}\omega_{j}\rho_{ij}\sigma_i\sigma_j, 
\end{align}
where $\rho_{ij}$ is the Pearson cross-correlation between $r_i$ and
$r_j$, and $\sigma_i$ and $\sigma_j$ are the standard deviations of the
return time series $r_i$ and $r_j$, respectively. The optimal portfolio
weights are determined by maximizing the portfolio return
$\Phi=\sum\limits_{t=1}^{T} \Pi(t)$ under the constraint that the risk
of the portfolio equals some fixed value $\Omega$. We formulate
maximizing $\Phi$ subject to these constraints as a quadratic
optimization problem
\begin{align}
\omega^{T}\Sigma\omega-q*\mathrm{R}^{T}\omega.
\end{align}
Here $\Sigma$ is the covariance matrix of the return time series for $m$
stocks. The parameter $q$ is the risk tolerance parameter with $q\in
(0,\infty)$. A large $q$ indicates investors have a strong tolerance to
risk and expect a large return. A small $q$ means investors are
extremely risk-adverse.  Optimal portfolios at different risk and return
levels can be displayed as an \textit{efficient frontier}, which is a
plot of the return $\Phi$ as a function of risk $\Omega$.

When determining the $m$ constitute stocks for a specific portfolio,
nodes inside a community indicate that connections within the community
are stronger than connections with nodes in other communities. Recent
research on the community detection of cross-correlation matrices found
that using the optimization procedure of modified modularity reveals
that residual cross-correlations among stocks from different communities
are negative \cite{MacMahon2015}. This qualitatively proves the
effectiveness of a community-based portfolio.

Here we use the community partitions of correlation-based networks for
different stock markets to construct our portfolio. In particular we use
stocks from different communities. When the community detection
algorithm shows that a PMFG network has $n_c$ communities, we randomly
choose one stock from each community. These stocks must only connect
with stocks within their own community, and this produces a portfolio
that consists of stocks that are weakly or even negatively
correlated. For the sake of comparison, we also construct a portfolio
made up of stocks residing in the same community. The stocks within the
same community are strongly correlated. Thus the risk diversification of
the portfolio with stocks from one community is relatively weak.

\begin{figure}[ht]
\centering
\includegraphics[width=\textwidth]{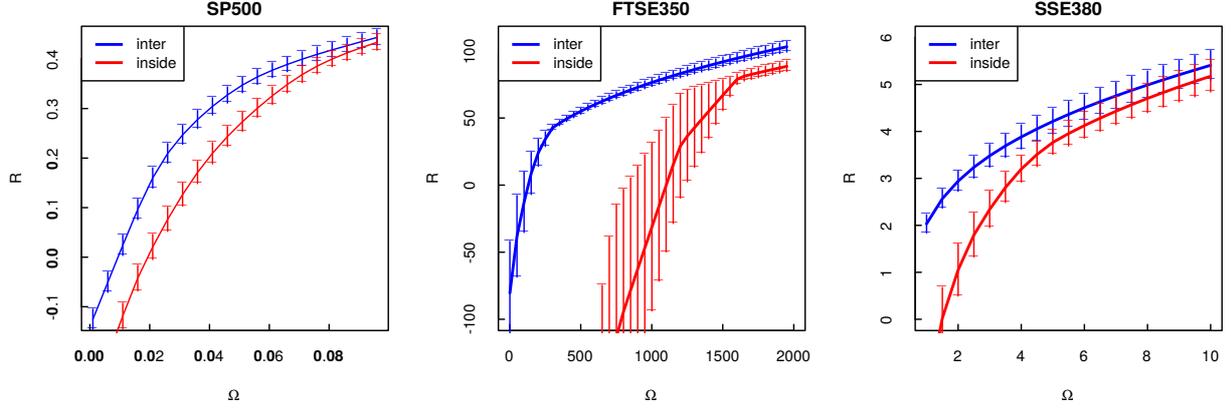}
\caption{\label{eff}The efficient frontiers for three different stock
  markets. The left, center and right columns are the results for S\&P
  500, FTSE 350 and SSE 380 respectively. The red lines are the results
  for these portfolios constructed by the stocks from the same
  community. The blue lines are the results for these portfolios
  constructed by the stocks from different community. The error bar
  represents the variance of return $R$ at each risk level
  $\Omega$. Here the portfolio size $m=30$. We have tested the portfolio
  size from $m=5$ up to $m=60$, the results are consistent.}
\end{figure}

Figure \ref{eff} shows the efficient frontiers of portfolios with stocks
from different communities---inter-portfolio (blue lines)---and of
portfolios with stocks from the same community---intra-portfolio (red
lines). Note that the inter-community portfolios have lower risk and
higher return in the three markets. Thus the risk diversification of the
inter-community portfolios are much stronger than that of
intra-community portfolios.

\subsubsection{Expected shortfall approach}

\noindent
The mean-variance portfolio framework assumes that normality of the return
 time series which is rarely valid in the real-world. Thus accurately
measuring portfolio risk is quite important. Except the variance as a risk
measure, the expected shortfall (ES) is a more modern coherent risk measure
\cite{Acerbi22001,Caccioli2013,CACCIOLI2016}. We designate $X$
to be the profit loss of a portfolio in the time range $(0, T)$, and
$\alpha = \eta\% \in (0, 1)$ to be a specified probability level. Then
the expected $\eta\%$ shortfall of a portfolio is
\begin{align}
{\rm ES}^{\alpha}(X)=-\frac{1}{\alpha}(E[X\textbf{1}_{X\leq
    x^{\alpha}}]-x^{\alpha}(\mathrm{P}[X\leq x^{\alpha}]-\alpha)). 
\end{align}
Here ES is the expected loss incurred in the $\eta\%$ worst cases of the
portfolio and is a coherent risk measure.  For a portfolio
$\{\omega_{i},i=1,\ldots,m\}$ of size $m$ stocks, when the return time
series is $\{r_i,i=1,\ldots ,m\}$, we minimize the ES$^{\alpha}$ of the
portfolio under the constraint of
$\sum\limits_{i=1}^{m}\omega_{i}=1$. We then set the probability level
$\alpha=95\%$ for the expected shortfall ES$^{\alpha}$ of the portfolio
and prohibit short selling.

\begin{figure}[ht]
\centering
\includegraphics[width=\textwidth]{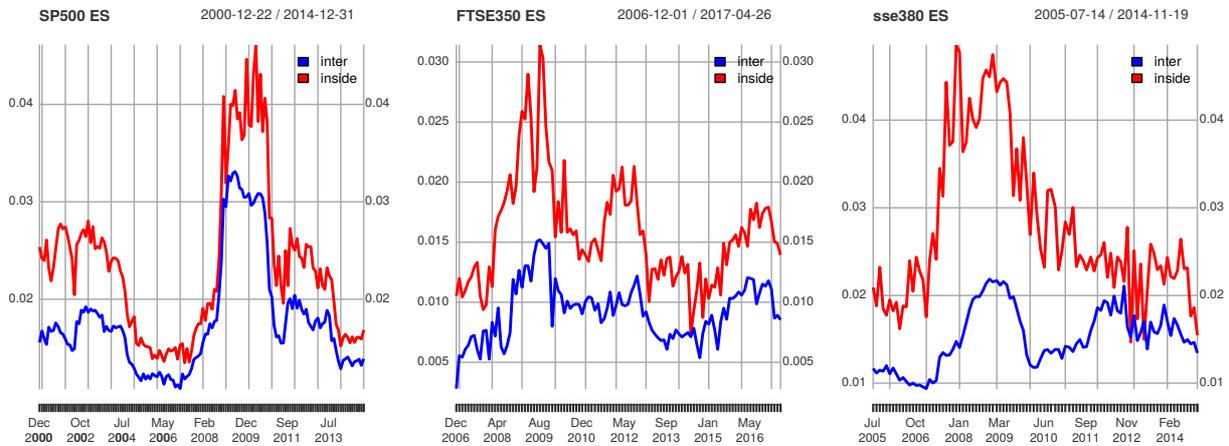}
\caption{\label{ES}The expected shortfall time series for three stock
  markets during the whole observation period. The left, center and
  right columns are the results for S\&P 500, FTSE 350 and SSE 380,
  respectively. The blue(red) lines are the expected shortfalls for the
  portfolios constructed by inter(inside) community stocks.}
\end{figure}

Figure\ref{ES} shows the time evolution of the expected shortfalls for
the three stock markets. The blue (red) lines represent the expected
shortfalls for inter- (intra-) portfolios. Note that the expected
shortfalls for the inter-community portfolios are much smaller than for
the intra-community portfolios. The monthly time series of expected
shortfalls for the three stock markets consistently quantifies the market
risk. We use the optimization procedure under the normalization
constraint to obtain the expected shortfall value. This means the
optimized risk in intra-community portfolios is much larger than that in
inter-community portfolios. Reference~\cite{MacMahon2015} proposes that
the correlation matrix is a weighted fully-connected network. There is a
negative correlation between the indices from different communities.
The lower expected shortfalls of inter-community portfolios have verified
this argument. Note that the ES value increases during crisis periods in all
three markets. The ES increases rapidly during market turbulence such as
the subprime crisis beginning in September 2008 and the European debit
crisis in the following year.

\begin{figure}[ht]
\centering
\includegraphics[width=\textwidth]{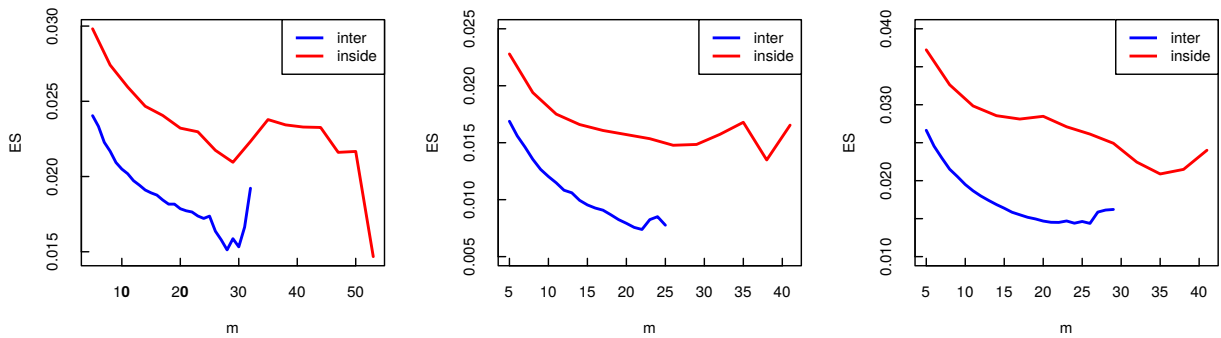}
\caption{\label{ES-out}The expected shortfalls for three stock markets
  with different portfolio sizes. The left, center and right columns are
  the results for S\&P 500, FTSE 350 and SSE 380, respectively. The
  blue(red) lines are the expected shortfalls for the portfolios
  constructed by inter(inside) community stocks.}
\end{figure}

Figure~\ref{ES-out} shows the dependency of ES on portfolio size
$m$. Note that the ES of inter-community portfolios are lower than those
of intra-community portfolios for different portfolio sizes $m$. The
largest inter-community portfolio size is smaller than the largest
intra-community portfolio. This is because the number of communities
yielded by the community detection algorithm is smaller than the size of
the largest community.
 
\section{Conclusion\label{conclusion}}

\noindent
We have analyzed the community structures of the correlation-based
networks of three stock markets. The community structures of the
correlation-based networks reveal clear modular stock market
structures. Using community structure information, we construct risk
diversified portfolios with higher return and lower risk. Using both
mean-variance and expected-shortfall frameworks, we construct portfolios
and find that the portfolios with inter-community stocks outperform these with
intra-community stocks. This is the first study that uses a community
detection algorithm to analyze the time evolution of correlation-based
networks. We apply community partition information to portfolio
selection and find that the community structure of the correlation-based
networks of stock markets is important.

\section{Acknowledgments}

\noindent
 This work is supported by Scientific and technological activities support funding for returned overseas students of Shaanxi province No. 27 and National Natural Science Foundation of China 71901171 and 72071006, Startup funding of NWPU G2021KY05101 and the interdisciplinary research fund for liberal arts of NWPU 21GH031109.
%\section*{References}
%\bibliographystyle{unsrt}
%\bibliography{/home/oliver/Work/ArticleAndReports/PhD_thesis/ZLF-CCNU/Biblio/library}

 \end{document}